\documentclass[prx,nofootinbib,twocolumn,showkeys,groupaddress,preprintnumbers,floatfix,superscriptaddress]{revtex4-1}

\usepackage{etex}                        \usepackage{ifpdf}                       \usepackage{xspace}                      \usepackage[table]{xcolor}               \usepackage{setspace}                    
\PassOptionsToPackage{final}{graphicx}   \usepackage{graphicx}                    
\makeatletter                            \@ifclassloaded{beamer}{}{\PassOptionsToPackage{pagebackref}{hyperref}}
\makeatother
\usepackage{hyperref}
\usepackage{url}                         \usepackage[super]{nth}                  \usepackage{contour}                     \usepackage{verbatim}                    \usepackage{ragged2e}                    \usepackage{attrib}                      \usepackage{adjustbox}                   \usepackage[absolute,overlay]{textpos}   \usepackage{calc}                        
\usepackage{amsmath}                     \usepackage{amscd}                       \usepackage{amsfonts}                    \usepackage{amssymb}                     \usepackage{amsthm}                      \usepackage{scalerel}                    \usepackage{bm}                          \usepackage{bbm}                         \usepackage{braket}				               \usepackage{mathtools}                   \usepackage{etoolbox}                    \usepackage{textcomp}                    \usepackage{stmaryrd}                    \usepackage{nicefrac}                    \usepackage[binary-units]{siunitx}       
\usepackage[T1]{fontenc}                 \usepackage{lmodern}                     \usepackage[scaled=0.72]{beramono}       \usepackage{microtype}                   
\usepackage{algorithm}                   \usepackage{algorithmicx}                
\usepackage{booktabs}                    \usepackage{dcolumn}                     
\makeatletter
  \@ifclassloaded{revtex4-1}
  {
    \usepackage[caption=false]{subfig}                                              }
  {
    \usepackage{caption}
    \usepackage{subcaption}
  }
  \@ifclassloaded{beamer}
  {
    \usepackage[backend=biber,
                autocite=superscript,
                doi=false,
                url=false,
                sorting=none]{biblatex}
    \usepackage[useregional,showdow]{datetime2}
  }
  {
        \usepackage{enumitem}                  }
\makeatother

\definecolor{ryan}{RGB}{64, 0, 64}
\definecolor{nix}{RGB}{255, 0, 0}

\definecolor{ucdblue1}{cmyk}{.87,.46,0,.49} \definecolor{ucdblue2}{cmyk}{1., .56, 0., .34}
\colorlet{ucdblue}{ucdblue2}

\makeatletter
\def\@lox@prtc{\section*{\@fxlistfixmename}\begingroup\def\@dotsep{4.5}}
\def\@lox@psttc{\endgroup}
\makeatother

\usepackage{epigraph}

\colorlet {past_color}    {red}
\colorlet {pres_color}    {blue}
\colorlet {futu_color}    {black!30!green}

\colorlet {temp_color_1}  {red!50!blue}
\colorlet {temp_color_2}  {red!50!green}
\colorlet {temp_color_3}  {blue!50!green}

\colorlet {hmu_color}     {blue!67!green}
\colorlet {rhomu_color}   {temp_color_1!80!blue}
\colorlet {rmu_color}     {blue}
\colorlet {bmu_1_color}   {temp_color_1}
\colorlet {bmu_2_color}   {temp_color_3}
\colorlet {qmu_color}     {temp_color_1!67!green}
\colorlet {wmu_color}     {temp_color_2!57!blue}
\colorlet {sigmamu_color} {temp_color_2}

\usepackage{listings}
\lstdefinestyle{mypython}{
language=Python,                        basicstyle=\small\ttfamily,             keywordstyle=\color{green!50!black},    commentstyle=\color{gray},              numbers=left,                           numberstyle=\tiny,                      stepnumber=1,                           numbersep=5pt,                          backgroundcolor=\color{gray!10},        frame=none,                             tabsize=2,                              captionpos=b,                           breaklines=true,                        breakatwhitespace=false,                showspaces=false,                       showtabs=false,                         morekeywords={as},                      }

\theoremstyle{plain}    
\theoremstyle{plain}    
\theoremstyle{plain}    
\theoremstyle{plain}    
\theoremstyle{plain}    
\theoremstyle{plain}    
\theoremstyle{plain}    
\theoremstyle{plain}    
\theoremstyle{plain}    
\theoremstyle{plain}    
\theoremstyle{plain}    
\theoremstyle{plain}

\newcommand{\CausalState}       { \mathcal{S} }

\newcommand{\forward}{+}
\newcommand{\reverse}{-}
\newcommand{\forwardreverse}{\pm} 

\newcommand{\FutureCausalState} { {\CausalState}^{\forward} }

\newcommand{\PastCausalState}   { {\CausalState}^{\reverse} }

\newcommand{\lastindex}[2]{
  \edef\tempa{0}
  \edef\tempb{#2}
  \ifx\tempa\tempb
        \edef\tempc{#1}
  \else
        \edef\tempa{0}
    \edef\tempb{#1}
    \ifx\tempa\tempb
      \edef\tempc{#2}
    \else
      \edef\tempc{#1+#2}
    \fi
  \fi
  \tempc
}

\newcommand{\CSjoint}[1][,]{
   \edef\tempa{:}
   \edef\tempb{#1}
   \ifx\tempa\tempb
            \ensuremath{\FutureCausalState\!#1\PastCausalState}
   \else
            \ensuremath{\FutureCausalState#1\PastCausalState}
   \fi
}
\newcommand{\CSjointKL}[3][,]{
   \edef\tempa{:}
   \edef\tempb{#1}
   \ifx\tempa\tempb
            \ensuremath{\FutureCausalState_{#2}\!#1\PastCausalState_{#3}}
   \else
            \ensuremath{\FutureCausalState_{#2}#1\PastCausalState_{#3}}
   \fi
}

\newif\ifpm
\edef\tempa{\forwardreverse}
\edef\tempb{\pm}
\ifx\tempa\tempb
   \pmfalse
\else
   \pmtrue
\fi

\ifcsname mathclap\endcsname
  \relax
\else
  \def\clap#1{\hbox to 0pt{\hss#1\hss}}

\fi

\newcommand{\op} [3] [] {
  \ensuremath{
    \operatorname{#2_{#1}}
    \if\relax\detokenize{#3}\relax
    \else
      \left[ #3 \right]
    \fi
  }
  \xspace
}

\makeatletter
  \@ifclassloaded{beamer}
  {
    \hypersetup{
      linkcolor={blue!50!black},
      citecolor={blue!50!black},
      urlcolor={blue!80!black},
    }
  }
  {
    \hypersetup{
      colorlinks,
      linkcolor={blue!50!black},
      citecolor={blue!50!black},
      urlcolor={blue!80!black},
    }
  }
\makeatother

\usepackage{empheq}
\usepackage{overpic}
\usepackage{qcircuit}

\newcommand{\argmin}{\text{argmin}}

\newcommand{\kB}{k_\text{B}}

\newcommand{\drive}{x_{0:\tau}}

\newcommand{\stationary}{\boldsymbol{\pi}}

\newcommand{\EP}[1][\rho_0]{\boldsymbol{\Sigma}_{#1}}
\newcommand{\EF}[1][\rho_0]{\boldsymbol{\Phi}_{#1}}  \newcommand{\tr}{\text{tr}}

\newcommand{\trB}{\tr_{\mathbb{B}}}
\newcommand{\trsys}{\tr_\text{sys}}
\newcommand{\trallbutb}{\tr_{\text{sys}, \mathbb{B} \setminus b}}

\newcommand{\reset}{r_\tau}

\newcommand{\q}{\alpha}

\begin{document}

\def\ourTitle{The impossibility of Landauer's bound for almost every quantum state
}

\def\ourAbstract{The thermodynamic cost of resetting an arbitrary initial state to a particular desired state
is lower bounded by Landauer's bound.  However, here we demonstrate that this lower bound is necessarily unachievable 
for nearly every initial state, for any reliable reset mechanism.
Since local heating threatens rapid decoherence, this issue is of substantial importance 
beyond mere energy efficiency.
For the case of qubit reset, we find the minimally dissipative state analytically for any reliable reset protocol,
in terms of the entropy-flow vector introduced here.
This allows us to verify a recent theorem about initial-state dependence of 
entropy production for any finite-time transformation, as it pertains to quantum state preparation.
}

\def\ourKeywords{  nonequilibrium thermodynamics, entropy
  production, relative entropy, open quantum systems
}

\hypersetup{
  pdfauthor={Paul M. Riechers},
  pdftitle={\ourTitle},
  pdfsubject={\ourAbstract},
  pdfkeywords={\ourKeywords},
  pdfproducer={},
  pdfcreator={}
}

\title{\ourTitle}

\author{Paul M. Riechers}
\email{pmriechers@gmail.com}

\affiliation{School of Physical and Mathematical Sciences, Nanyang Technological University, 
637371 Singapore}

\affiliation{Complexity Institute, Nanyang Technological University, 
637335 Singapore}

\author{Mile Gu}
\email{ceptryn@gmail.com}

\affiliation{School of Physical and Mathematical Sciences, Nanyang Technological University, 
637371 Singapore}

\affiliation{Complexity Institute, Nanyang Technological University, 
637335 Singapore}

\affiliation{Centre for Quantum Technologies, National University of Singapore, 
3 Science Drive 2,
117543 Singapore}

\date{\today}
\bibliographystyle{unsrt}

\begin{abstract}
\ourAbstract
\end{abstract}


\date{\today}
\maketitle

\setstretch{1.1}

\section{Introduction}

Whether initializing a quantum computer or 
a quantum experiment,
a desired quantum state must be prepared.  The thermodynamic cost of erasing the pre-existing
state and replacing it with a 
newly prepared state is typically associated with Landauer's bound---the accepted
exchange rate
between information and energy~\cite{Land61a, Parr15}.
Indeed,
Landauer's bound is one of the key results tying quantum information to physical predictions, 
via the universality of thermodynamics.  

In the simplest quantum version of this bound,
the expected heat $Q$ released to the environment 
must exceed the reduction in von Neumann entropy of the system 
(multiplied by the thermal energy $\kB T$ of the environment)~\cite{Alic04, Reeb14}.
It is generally accepted that in the limit of quasistatically
slow transformations,
the heat is expected to approach Landauer's lower bound~\cite{Jun14a, Mill20}.

However, 
in the following,
we demonstrate that no reliable protocol for preparing a quantum state
can achieve Landauer's bound 
for any more than, at best, one of infinitely many possible inputs to the preparation device.
In particular, when resetting a quantum state via any reliable mechanism, 
we find that 
there is only a single input state $\alpha_0$ 
leading to minimal entropy production
(which is generically a mixed state)
among the uncountably infinite number of possible input states.

Suppose, for example, that a quantum state is reset 
in finite time 
in an environment of ambient temperature $T$,
and assume the validity of the Second Law of thermodynamics.
Then---even if the reset mechanism approaches Landauer's bound for the minimally dissipative input---the reset 
will necessarily produce more than Landauer's required heat $Q_\text{Landauer}$ for every other input.
In particular, 
for any input $\rho_0$ to any reliable erasure protocol,
\begin{align}
Q - Q_\text{Landauer} \geq \kB T \, \text{D}[ \rho_0 \| \q_0] ~,
\label{eq:MotivatingEq}
\end{align}
where 
$\text{D}[ \rho_0 \| \q_0]$ is the quantum relative entropy between the actual input $\rho_0$
and the minimally dissipative input $\alpha_0$.

In the following, we demonstrate the predictive power of this result
and unravel its important implications for quantum state preparation.
To get there, 
\S \ref{sec:Background} first reviews and further develops the relevant theory,
leading to a generalization of Eq.~\eqref{eq:MotivatingEq},
which applies to finite-time transformations of any physical system placed in 
any environment.
\S \ref{sec:Obs} points out the immediate consequences.
To address the thermodynamic
cost of initializing any quantum computing algorithm,
we then develop further analytic results for qubit erasure 
in \S \ref{sec:q0_for_qubit_erasure}.
We find the exact minimally dissipative quantum state analytically 
for any reliable qubit-reset protocol,
in terms of the \emph{entropy-flow vector} introduced here.
The entropy flow vector, in turn, is found algebraically for any qubit transformation protocol,
via experimentally obtained heats, from any four linearly independent initial states.
In \S \ref{sec:Examples}, 
we demonstrate our results with explicit physical models for qubit erasure.
This allows us to verify the recent theoretical results for initial-state dependence of entropy production,
developed
for any finite-time transformation in Ref.~\cite{Riec20a},
as applied to quantum state preparation.
Finally, we examine the thermodynamic penalty of decoherence,
and show that this bounded contribution from decoherence is often overshadowed by 
the more dire penalty imposed by the purity of the minimally dissipative input state.

The unachievability of Landauer's bound has been explored from several other directions before.
For example, 
Ref.~\cite{Reeb14} showed that finite-dimensional quantum baths can never exactly achieve Landauer's bound,
although Landauer's bound can be approached in the limit of large baths.
And, in the single-shot regime, the IID limit of many copies of the system is necessary to 
reliably approach Landauer's bound in a single shot (rather than just on average)~\cite{Dahl11, Halp15}.
However, our result is even more severe than these previously acknowledged limitations of Landauer's bound.
Even in the case of large baths, and when concerned with only average quantities, we find that
Landauer's bound still cannot be approached by more than a single initial quantum state 
for any reliable implementation of reset.

\section{Background: Thermodynamics of Quantum Reset}
\label{sec:Background}

Nonequilibrium thermodynamics is largely a theory of entropy production.
For example, the Second Law of thermodynamics states that 
entropy production is expected to be non-negative 
for any thermodynamic process, from any initial state $\rho_0$:
\begin{align}
\EP \geq 0 ~.
\label{eq:2ndLaw}
\end{align}
Validity of the Second Law requires that the system is initially uncorrelated with the environment,
and each part of the environment is initially uncorrelated and in local equilibrium~\cite{Espo10a, Reeb14, Riec20a}. 
This means that the initial joint state of system and environment is a product state $\rho_0^\text{tot} = \rho_0 \otimes \rho_0^\text{env}$  
with
$\rho_0^\text{env} = \bigotimes_{b \in \mathbb{B}} \stationary^{(b)}$,
where $\stationary^{(b)}$ is a state of local equilibrium.
However, correlation between system and baths, as well as correlation among the baths,
generically builds up over time, while the baths also generically depart from local equilibrium~\cite{Ptas19a}.

In the quantum setting,
the expected entropy production (from time $0$ to $\tau$) 
is the \emph{expected entropy flow} $\EF$ to the environment
beyond any compensating reduction in the von Neumann entropy of the system $S(\rho_t) = - \tr( \rho_t \ln \rho_t)$:
\begin{align}
\EP = \EF + \kB S(\rho_\tau) - \kB S(\rho_0) ~,  \label{eq:EPdef}
\end{align}
where $\kB$ is Boltzmann's constant.

App.~\ref{sec:EF_def} provides the most general definition of the expected entropy flow, 
which gives the most general form of our results.
However, to demonstrate the main results, it should suffice to consider the case 
that the environment is composed of a set of baths $\mathbb{B}$, 
each initially in either canonical or grand canonical equilibrium.
If each bath has an initial temperature $T^{(b)}$ and 
chemical potentials $\{ \mu^{(b, \ell)} \}_\ell$,
then 
the expected entropy flow takes on 
the 
familiar form~\cite{deGr84, Kond14, Espo10a, Ptas19a}:
\begin{align}
\EF
= \sum_{b \in \mathbb{B}} \frac{Q^{(b)}}{T^{(b)}} - \frac{1}{T^{(b)}} \sum_{\ell} \mu^{(b, \ell)} \Delta \braket{N^{(b, \ell)}} ~,
\label{eq:SenvDef1}
\end{align}
where the heat $Q^{(b)} = \Delta \tr(\rho_t^{(b)} H^{(b)})$ 
is the expected energy change of
bath $b$
over the course of the process and 
$\Delta \braket{N^{(b, \ell)}} = \Delta \tr(\rho_t^{(b)} N^{(b, \ell)}) $ is the expected change in the bath's number of $\ell$-type particles.
Here, $\rho_t^{(b)}$ is the reduced state of bath $b$ at time $t$, 
$H^{(b)}$ is the bath's Hamiltonian,
and $N^{(b, \ell)}$ is one of its number operators.

Eqs.~\eqref{eq:2ndLaw} and \eqref{eq:EPdef} together immediately imply 
a very general form of Landauer's bound:
\begin{align}
\EF \geq  \kB S(\rho_0) - \kB S(\rho_\tau) ~.
\label{eq:GenLand}
\end{align}
In other words:
Whenever the Second Law of thermodynamics is valid,
the change in the system's entropy bounds the expected entropy flow.

If there is a single thermal bath at temperature $T$, then the expected entropy flow is simply 
$$\EF = Q / T ~,$$
where $Q$ denotes the expected heat flow to the bath---i.e., its change in energy.
Landauer's bound for the heat released to the environment during reset 
is then $Q \geq Q_\text{Landauer}$ with 
$Q_\text{Landauer} \equiv \kB T [  S(\rho_0) - S(\rho_\tau)  ]$.
Paradigmatic `bit erasure' 
takes a completely mixed state $\rho_0 = I/2$ 
to a pure state $\rho_\tau = \ket{0} \! \bra{0}$, yielding the $S(\rho_0) - S(\rho_\tau) = \ln 2$
that leads to the most familiar form of Landauer's bound
$$Q \geq \kB T \ln 2$$ 
for the heat required to erase either a bit or a qubit.

Entropy production quantifies the heat (or entropy flow, more generally)
produced beyond Landauer's bound.  
Positive entropy production implies thermodynamic irreversibility---an effectively 
irreversible loss of thermodynamic resources.
The cause of this effective irreversibility is easier to understand 
when we decompose the
expected entropy production
into a collection of non-negative contributions, as in Ref.~\cite{Espo10a}.
The decomposition
shows that the non-negativity of entropy production can be attributed to both 
1)
the growth of total correlation
among the system and all baths and
2)
the nonequilibrium addition to free energy built up in each bath.
Both of these structural features are assumed to be too difficult to practically leverage,
which is the reason for the effective loss of useful resources.

The thermodynamic process can be made explicit through the control protocol $\drive$
which implies the trajectory of the joint system--baths Hamiltonian $H_{x_t}^\text{tot}$ 
from time $0$ through $\tau$.
The control protocol induces a net time evolution $\mathcal{U}_{\drive}$ 
of the system--baths mega-system,
so that the system's state at the end of the transformation 
is
\begin{align}
\rho_{\tau} &= \Gamma(\rho_0) 
= \trB \bigl( \mathcal{U}_{\drive} \rho_0 \otimes \rho_0^\text{env}
\mathcal{U}_{\drive}^\dagger \bigr) ~.
\label{eq:NetUnitaryEvolution}
\end{align}
In general, $x_t$ may consist of any number of parameters.
Different thermodynamic processes correspond to different control protocols.
In the following, we consider \emph{reliable reset} protocols, which induce the net transformation
\begin{align}
\Gamma(\rho_0) \approx \reset
\end{align}
to the input-independent state $\reset$.
The reliability can be quantified by the maximal trace distance between the final and desired state,
$\epsilon = \sup_{\rho_0} \bigl\| \Gamma(\rho_0) - \reset \bigr\|_1$, which should be very nearly zero. 
The following results therefore apply
not only to erasing a qubit but also
to many other important scenarios, including:
1)
to initialize any number of registers of a quantum computer,
2)
to produce a maximally mixed state,
3)
to achieve equilibrium,
4)
to establish a nonequilibrium steady state,
and
5)
to prepare a special state like a Bell state.

For any thermodynamic process, the expected entropy production from time $0$ through $\tau$
will depend on the initial state of the system.
It is useful to consider any minimally dissipative state
\begin{align}
\q_0
\in
\argmin_{\rho_0} \EP ~.
\end{align}
Ref.~\cite{Riec20a} established that  
the expected entropy production from any initial state $\rho_0$ 
that acts on the support of $\q_0$
is quantified by:
\begin{align}
\EP - \EP[\alpha_0] = \kB \text{D} [ \rho_0 \| \alpha_0 ] - \kB \text{D} [ \rho_\tau \| \alpha_\tau ] ~,
\label{eq:QKW_result}
\end{align}
where $\text{D}[\rho \| \q] = \tr(\rho \ln \rho) - \tr(\rho \ln \q)$ is the quantum relative entropy.
This generalizes the classical result of Ref.~\cite{Kolc17},
where the quantum relative entropy between quantum states
reduces to the Kullback--Leibler divergence between classical distributions.
App.~\ref{sec:EP_InitStateDependence}
further extends the generality of this result, 
proving that it remains valid even if thermodynamic baths have time-varying properties.
Eq.~\eqref{eq:QKW_result} is very general and is valid 
in the presence of arbitrary initial environments, even when the Second Law is not.
(The examples to follow, however, all fall within the purview of the Second Law.)

In the $\epsilon \to 0$ limit of high-fidelity reset,
Ref.~\cite{Riec20a} showed that 
$\text{D} [ \rho_\tau \| \alpha_\tau ] \to 0$ and so
the difference in expected entropy production
from any other initial density matrix $\rho_0$
is exactly proportional to the initial distinguishability between $\rho_0$ and $\q_0$ as quantified by the quantum relative entropy:
\begin{align}
\EP = \EP[\q_0] + \kB \text{D}[\rho_0 \| \q_0] ~.
\label{eq:MainResetEqForEP}
\end{align}
In the following, we will demonstrate the validity of this result via several explicit examples.

The next section explicates several important implications of Eq.~\eqref{eq:MainResetEqForEP}. 
Subsequently,
Sec.~\ref{sec:q0_for_qubit_erasure}
shows how the minimally dissipative state can be found
from the entropy flow vector, which will be introduced shortly.
This transforms Eq.~\eqref{eq:MainResetEqForEP} 
from a neat theoretical result 
into a powerful predictive tool for real physical systems.

\section{Observations}
\label{sec:Obs}

Our first observation is that 
\emph{there is always a unique 
minimally dissipative initial state for any 
reliable reset protocol}.
If $\q_0$ has full support, this follows from Eq.~\eqref{eq:MainResetEqForEP}
and the fact that the 
quantum relative entropy $\text{D}[\rho_0 \| \q_0]$ is positive unless $\rho_0 = \q_0$. 
In fact, App.~\ref{sec:q0HasFullSupport} guarantees that $\q_0$ has full support
on the domain of states to be reset,
which proves the observation.

The dissipation from most inputs grows drastically 
as a reset protocol is adjusted to bring the minimally dissipative state closer to a pure state.
Indeed, via a well known sensitivity of the relative entropy,
the dissipation diverges for most inputs as $\q_0$ approaches a pure state.
This will be demonstrated in the examples of \S \ref{sec:Examples}.

To avoid this divergent dissipation,
it may be desirable to design a reset protocol to be thermodynamically optimal for the completely mixed state $\q_0 = I/d$
of the $d$-dimensional quantum system.
What is the heat required to reliably reset any qudit to a pure state,
when the system is coupled to a single thermal bath at temperature $T$ and $\EP[I/d] = 0$?  From Eq.~\eqref{eq:MainResetEqForEP},
we observe
that 
\emph{the heat in this case is}
$$Q = \kB T  \text{D} \bigl[ \rho_0 \big\| I / d \bigr] + \kB T S(\rho_0)  = \kB T \ln d ~,$$
\emph{independent of the initial state}.
While $\kB T \ln d$ may be reminiscent of Landauer's bound,
it should be noted that this exceeds Landauer's bound of $\kB T S(\rho_0)$
by $\kB T  \text{D}[ \rho_0 \| I / d ]$.
If the input is not completely mixed, then less heat is possible---by designing the reset protocol to be optimal for the true input.
However, when the true input is unknown, 
or if a single protocol is part of a prefabricated design meant to reset states of diverse origin,
this may be an acceptable price for ignorance.

Further general insight can be gained by decomposing the quantum relative entropy into
contributions from
classical relative entropy and quantum coherence.
The classical relative entropy  
involves two classical probability distributions
$\mathcal{Q}_0$ and $\mathcal{P}_0$ over a complete orthonormal set of $\alpha_0$'s right 
eigenstates.
Specifically, 
$\mathcal{Q}_0(s) = \braket{s | \q_0 | s }$ is the probability that $\q_0$ 
would be found in the eigenstate $\ket{s}$
if the state is projectively measured in its eigenbasis.
Similarly, $\mathcal{P}_0(s) = \braket{s | \rho_0 | s}$ is the probability that $\rho_0$ would appear 
to be 
in the state $\ket{s}$
if measured in the eigenbasis of $\q_0$.

It is useful to note that the quantum relative entropy can always be decomposed as~\cite{Riec20a}
\begin{align}
\text{D}[\rho_0 \| \q_0] 
&=
 \text{D}_\text{KL} \bigl[ \mathcal{P}_0 \big\| \mathcal{Q}_0 \bigr] +
 C_{\q_0}(\rho_0) 
 ~.
\label{eq:RelEnt_asDKLplusCoherence}
\end{align}
Above,
$ \text{D}_\text{KL} \bigl[ \mathcal{P}_0 \big\| \mathcal{Q}_0 \bigr] $ 
is the classical relative entropy, also known as the Kullback--Leibler divergence, 
which is always non-negative.
Finally,
$ C_{\q_0}(\rho_0)$ 
is the \emph{relative entropy of coherence}~\cite{Baum14},
which quantifies the coherence of $\rho_0$ on the eigenbasis of $\q_0$.
\footnote{The relative entropy of coherence can be written as 
$C_{\q_0}(\rho_0) = 
S( P_0 ) - S(\rho_0)$
where $P_0 = \sum_{s \in \mathbb{E}_{\q_0}} \mathcal{P}_0(s) \ket{s} \! \bra{s}$
is the initial state decohered in $\q_0$'s eigenbasis $\mathbb{E}_{\q_0}$.}
The relative entropy of coherence is also non-negative. 
The non-negativity of the two terms in this decomposition 
leads to our 
third observation, that 
\emph{any initial coherence on the minimally dissipative eigenbasis 
directly results in extra dissipation
during state reset, 
compared to the same state decohered in that basis}.

At the same time, we note that the 
the relative entropy of coherence is always upper bounded by a constant, $\ln d$, 
for a $d$-dimensional system.
Hence, 
for any process with very large entropy production $\EP - \EP[\q_0] \gg \ln d$,
the classical limit (i.e., ignoring the possibility of coherence) 
will suffice to explain the bulk of the dissipation.
In such cases, the Kullback--Leibler divergence is responsible for most of the dissipation,
with coherence playing a relatively minor role.

With the general theoretical framework laid out,
we can now circle back to our opening comments.
Whenever the Second Law is valid, the non-negativity of $\EP[\alpha_0]$ 
together with Eq.~\eqref{eq:MainResetEqForEP}
imply that 
\begin{align}
\EP \geq \kB \text{D}[\rho_0 \| \q_0] 
\label{eq:GenIntroInequality}
\end{align} 
for any reliable reset protocol.
This generalization of Eq.~\eqref{eq:MotivatingEq} is 
necessary when heat does not capture all aspects of entropy flow.
It should be noted, however, that Eq.~\eqref{eq:MainResetEqForEP}
is much more predictive than Eq.~\eqref{eq:GenIntroInequality},
since the former is an \emph{equality} in the limit of reliable reset.

In the next section, we introduce the entropy-flow vector,
and identify the minimally dissipative state for any
process that reliably resets a qubit.
This then allows us to draw further lessons from the investigation of several examples.

\section{The minimally dissipative state for qubit erasure}
\label{sec:q0_for_qubit_erasure}

Landauer's bound is typically associated with reset of a bit or qubit to a particular computational state $\texttt{0}$.
In the quantum setting, this corresponds to resetting the quantum state to a 
pure state $\reset = \ket{0} \! \bra{0}$.

The instantaneous state of a qubit 
can be fully described by its Bloch vector $\vec{a} = (a_x, a_y, a_z) \in \{ \vec{r} \in \mathbb{R}^3: |\vec{r}| \leq 1 \}$
via
$\rho_t = \tfrac{1}{2}(I + \vec{a}(t) \cdot \vec{\sigma})$,
where $\vec{\sigma} = (\sigma_x, \sigma_y, \sigma_z)$ is the vector of Pauli matrices. 

\subsection{The entropy-flow vector}

App.~\ref{sec:ExistenceAndConstruction_for_EFvec} shows that 
there is an input-independent \emph{entropy-flow vector} 
$\vec{\phi} = (\phi_x, \, \phi_y, \, \phi_z)$ for each 
control protocol
such that 
\begin{align}
\EF = \EF[I/2] +  \tfrac{1}{2} \vec{a} \cdot \vec{\phi}
\label{eq:EF_simpleLinAlg}
\end{align}
for any initial state $\rho_0 = I/2 + \tfrac{1}{2} \vec{a} \cdot \vec{\sigma}$.
Conveniently, 
\emph{the expected entropy flow from the completely mixed state $\EF[I/2]$ together with 
the entropy-flow vector $\vec{\phi}$ can both be inferred via simple linear algebra
when the expected entropy flow is measured from any four experimentally accessible initial states}.

Suppose we have access to any four linearly independent initial density matrices 
$\rho_0^{(n)}$ 
with Bloch vectors $\vec{a}^{(n)}$,
and their resultant
expected entropy flow $\EF[\rho_0^{(n)}]$, for $n \in \{1, 2, 3, 4 \}$. 
Almost any four initial density matrices chosen at random would suffice, 
since it is unlikely that a random density matrix will lie in the subspace
spanned by the previously chosen density matrices.
As a concrete and likely useful example,
we could choose the four initial states to be the completely mixed state $I/2$,
together with the pure $\hat{x}$, $\hat{y}$, and $\hat{z}$ states:
$(I + \sigma_x)/2$,  $(I + \sigma_y)/2$,  and $(I + \sigma_z)/2$,  respectively.

From Eq.~\eqref{eq:EF_simpleLinAlg},
we have
\begin{align}
\underbrace{
\begin{bmatrix}
2 & a_x^{(1)} & a_y^{(1)} & a_z^{(1)} \\
2 & a_x^{(2)} & a_y^{(2)} & a_z^{(2)} \\
2 & a_x^{(3)} & a_y^{(3)} & a_z^{(3)} \\
2 & a_x^{(4)} & a_y^{(4)} & a_z^{(4)}
\end{bmatrix}
}_{\equiv A}
\begin{bmatrix}
\EF[I/2] \\ 
\phi_x \\
\phi_y \\
\phi_z
\end{bmatrix}
=
2
\begin{bmatrix}
\EF[\rho_0^{(1)}] \\ 
\EF[\rho_0^{(2)}] \\
\EF[\rho_0^{(3)}] \\
\EF[\rho_0^{(4)}]
\end{bmatrix}
~.
\label{eq:LinAlgEF}
\end{align}
From Eq.~\eqref{eq:LinAlgEF},
we can obtain an expression for the entropy-flow vector:
\begin{align}
\begin{bmatrix} \EF[I/2] \\ \, \\ \vec{\phi} \\ \, \end{bmatrix}
= 2 \, A^{-1} \,  
\begin{bmatrix}
\EF[\rho_0^{(1)}] \\ 
\EF[\rho_0^{(2)}] \\
\EF[\rho_0^{(3)}] \\
\EF[\rho_0^{(4)}]
\end{bmatrix}
~.
\label{eq:EFvector_solution}
\end{align}
Note that the invertibility of $A$ has been assured by the linear independence of the four initial states.

Once $\EF[I/2]$ and $\vec{\phi}$
have been obtained via Eq.~\eqref{eq:EFvector_solution},
the entropy flow can be obtained analytically (for any thermodynamic process acting on a qubit)
for any initial state via Eq.~\eqref{eq:EF_simpleLinAlg}.
For any reliable erasure protocol,
this also leads to an analytic expression for entropy production from any initial state.

\subsection{Analytic entropy production}

For any reliable reset protocol,
the entropy of the final state is independent of the initial state.
In this low-$\epsilon$ limit where $S(\rho_\tau) = S(r_\tau)$,
Eq.~\eqref{eq:EPdef} tells us that entropy production can be expressed as
\begin{align}
\EP = \EF  + \kB S(r_\tau) - \kB S(\rho_0) ~.  
\label{eq:EPdef_forReliableReset}
\end{align}
When resetting to a pure state,
the entropy of the final state vanishes
and Eq.~\eqref{eq:EPdef_forReliableReset} further simplifies to
\begin{align}
\EP = \EF  - \kB S(\rho_0) ~.  
\label{eq:EPdef_forReliableErasure}
\end{align}
Moreover, the entropy of a qubit only depends on its purity, via the length of its Bloch vector $a = | \vec{a} |$.
This is because a qubit state always spectrally decomposes into 
$(1+a)/2$ times the pure state in the direction of the Bloch vector,
plus $(1-a)/2$ times the pure state in the antipodal direction on the Bloch sphere.
For an initial state with Bloch vector $\vec{a}$, the initial entropy is therefore
\begin{align}
S(\rho_0) = -\tfrac{1+a}{2} \ln \bigl(\tfrac{1+a}{2} \bigr) - \tfrac{1-a}{2} \ln \bigl( \tfrac{1-a}{2} \bigr) ~.
\label{eq:QubitEntropy}
\end{align}
Eqs.~\eqref{eq:EF_simpleLinAlg}, \eqref{eq:EPdef_forReliableErasure}, and \eqref{eq:QubitEntropy}
thus yield an analytic expression for entropy production 
of any reliable qubit erasure protocol
via 
1) 
the Bloch vector $\vec{a}$ of the initial state,
2)
the expected entropy flow $\EF[I/2]$ from the completely mixed state,
and 
3)
the entropy-flow vector $\vec{\phi}$ induced by the protocol.

\subsection{The minimally dissipative state}

We can now minimize Eq.~\eqref{eq:EPdef_forReliableReset} to find the minimally dissipative initial state $\q_0$ 
analytically in terms of the entropy-flow vector
for any reliable reset protocol.

We will parametrize a generic initial density matrix $\rho_0$ via
its Bloch vector $\vec{a} = (a_x, a_y, a_z)$ with magnitude
$a = (a_x^2 + a_y^2 + a_z^2)^{1/2}$.
The minimally dissipative initial state $\q_0$ has the 
Bloch vector $\vec{a}^*$.
Differentiating the entropy production with respect to changes in the Bloch vector, 
we find
\begin{align}
\frac{\partial \EP}{\partial a_x} = \frac{a_x}{2 a} \ln \bigl( \tfrac{1+a}{1-a} \bigr)  + \tfrac{1}{2} \phi_x
\end{align}
and similar expressions when we take the derivative with respect to $a_y$ or $a_z$.
Since $\q_0$ is a unique non-pure state,
the condition of minimization requires that 
\begin{align}
\frac{\partial \EP}{\partial a_x} \Bigr|_{a_x = a_x^*} = 0 
\end{align}
etc.\ for $a_y^*$ and $a_z^*$.
This immediately leads to conditions like
\begin{align}
a_x^* = - \phi_x \frac{a^*}{\ln \bigl( \tfrac{1+a^*}{1-a^*} \bigr)  }  
~.
\label{eq:First_axstar_condition}
\end{align}
Combining Eq.~\eqref{eq:First_axstar_condition} with the corresponding expressions for 
$a_y^*$ and $a_z^*$ according to $(a_x^{*2} + a_y^{*2} + a_z^{*2})^{1/2} = a^*$
yields a condition for $a^*$:
\begin{align}
\ln \bigl( \tfrac{1+a^*}{1-a^*} \bigr) = 
\sqrt{ \phi_x^2 +  \phi_y^2 +  \phi_z^2 } =  \phi
~,
\label{eq:First_astar_condition}
\end{align}
which can be solved to obtain
\begin{align}
a^* = 
\tanh(\phi/2)
~.
\label{eq:Final_astar_condition}
\end{align}
Plugging this back into Eq.~\eqref{eq:First_axstar_condition},
we find that
\begin{align}
\vec{a}^* = - \tanh(\phi/2) \, \hat{\phi} 
~,
\label{eq:Final_avec_condition}
\end{align}
where $\hat{\phi} =  \vec{\phi} / \phi $.
This gives the Bloch vector $\vec{a}^*$ of
the minimally dissipative initial state $\q_0$,
analytically
for any reliable reset protocol, 
in terms of the protocol's entropy-flow vector $\vec{\phi}$. 

Qualitatively, we can note from Eq.~\eqref{eq:Final_avec_condition} that the minimally
dissipative Bloch vector $\vec{a}^*$ points in the opposite direction of the entropy-flow vector $\vec{\phi}$.
This enables a reduction in heat flow.
As the magnitude of the entropy-flow vector $\phi$ grows beyond two,
the minimally dissipative Bloch vector
converges exponentially
to the edge of the Bloch sphere.
However, the minimally dissipative initial state is never exactly pure for any finite $\phi$,
since entropy production balances entropy flow against the growth in state uncertainty---and,
evidently,
a sufficiently small reduction in state uncertainty can always outweigh
the smaller potential reduction in entropy flow.

\section{Explicit Reset Protocols}
\label{sec:Examples}

\subsection{A first example: Reset via SWAP}

It should be noted that our preceding results are very generally applicable
and are not limited to weak interactions nor any of the other approximations that are required of the common 
quantum master equations often employed in quantum thermodynamics (although our results indeed apply there also).
To demonstrate our result in a setting of strong interactions with an explicit finite bath,
we first consider reset via a swap operation 
between the system and a part of an effectively cold bath.

This simple scenario corresponds to `throwing away' the old state $\rho_0$
and replacing it with a bath state $\gamma \approx \ket{0} \! \bra{0}$.
We consider a bath of $N$ independent qubits, $\stationary^{(b)} = \gamma^{\otimes N}$, which are initially in canonical equilibrium at temperature $T$, such that $\gamma = e^{-\beta H_b}/Z_b$.  
This bath of Gibbs states 
is a bath of nearly pure $\ket{0} \! \bra{0}$ states if the bath Hamiltonian for each qubit is $H_b = -E_b \sigma_z$
with the `effectively cold' condition that $E_b \gg \kB T$. 
To reset the state of the system,
we merely swap it with a state from the effectively cold bath.  This swap operation is implemented via a unitary operation on the joint system--baths megasystem, as depicted in the following quantum-circuit diagram:

\begin{align*}
\scalebox{1.0}{
	\Qcircuit @C=1em @R=.7em {
		\lstick{\gamma} & \qw & \qw & \qw & \rstick{\gamma} \\
		\lstick{\gamma} & \qw & \qw & \qw & \rstick{\gamma} \\
		\, & \, & \vdots \\
		\, & \, & \, \\
		\lstick{\gamma} & \qw & \qw & \qw & \rstick{\gamma} \\
		\lstick{\gamma} & \qw & \qswap & \qw & \rstick{\rho_0} \\
		\lstick{\rho_0} & \qw & \qswap \qwx[-1] & \qw & \rstick{\gamma} 
	}
}
\end{align*}			

Note that entropy flow, as defined in Eq.~\eqref{eq:SenvDef1},
can be quantified exactly via the \emph{initial} temperature of the bath, as in Refs.~\cite{Espo10a, Ptas19a}, for arbitrarily small environments.
All thermodynamic quantities can be calculated exactly: heat, change in von Neumann entropy, and entropy production.
Our calculation is valid for any 
Hamiltonian of the system and for any initial state $\rho_0$.
\footnote{
When the bath has many similar subsystems, 
the swap operation 
has a negligible effect on the overall temperature of the bath.
However, Eq.~\eqref{eq:SenvDef1} only requires the initial temperature to be well defined, 
and so does not require the large-$N$ limit.
}

The heat transferred to the bath is
\begin{align}
Q
&= 
\tr(\rho_\tau^{(b)} H^{(b)}) - \tr(\rho_0^{(b)} H^{(b)}) \nonumber \\
&= 
\tr(\rho_0 H_b) - \tr(\gamma H_b) \\
&= 
\kB T \tr(\gamma \ln \gamma) - \kB T \tr(\rho_0 \ln \gamma)
~,
\end{align}
where we have used the fact that 
$H_b = -\kB T \ln \gamma - \kB T \ln Z_b$.
Meanwhile, the swap operation changes the von Neumann entropy of the system according to 
\begin{align}
\Delta S_\text{sys} = \kB \tr(\rho_0 \ln \rho_0) - \kB \tr(\gamma \ln \gamma) ~.
\end{align}

Clearly, this reset-via-swap protocol has
a unique minimally dissipative initial state of $\q_0 = \gamma$, 
since then $Q=0$ and $\Delta S_\text{sys} = 0$ and so $\EP[\q_0] = 0$.
For any initial state $\rho_0$,
the entropy production will be:
\begin{align}
\EP 
&= 
Q / T + \Delta S_\text{sys}  \\
&= 
\kB \text{D} [ \rho_0 \| \gamma ] 
\label{eq:SwapBathRelEnt} 
~.
\end{align}

Clearly, Eq.~\eqref{eq:SwapBathRelEnt}
agrees with Eq.~\eqref{eq:MainResetEqForEP},  which is expected since Eq.~\eqref{eq:MainResetEqForEP}
quantifies the entropy production for reliable reset via \emph{any} means. 
This and the following examples
validate the more general result,
but also give us the opportunity to 
explore the more nuanced implications.

Notably, if $\gamma \approx \ket{0} \! \bra{0}$, then the entropy production diverges as $\rho_0 \to \ket{1} \! \bra{1}$.
This points to a disadvantage of 
the throw-away strategy to reset.
It can be \emph{thermodynamically advantageous to recycle} the system state,
as demonstrated in the next example 
where $\q_0$ can be made closer to the identity than to the desired final state.

It may be recognized that Eq.~\eqref{eq:SwapBathRelEnt}
has the form of a nonequilibrium addition to free energy.  
However, it is not related to the free energy of the system but, rather, is related to the change in nonequilibrium free energy \emph{of the bath},
which is consistent with 
the aforementioned decomposition of entropy production.
Note that 
Eq.~\eqref{eq:SwapBathRelEnt}
applies independently of the Hamiltonian of the system, so long as $\rho_0$ is the state of the system at the time of the swap.

\subsection{Reliable Reset Protocols via Time-dependent Lindbladians}

We now consider the scenario of single-qubit erasure
in the experimentally common regime 
well described by Markovian master equations.
Following Ref.~\cite{Mill20}, we
consider a family of protocols 
that utilize two time-dependent control parameters $x_t = (E_t, \theta_t)$,
which determine the 
time-varying Hamiltonian of the system:
\begin{align}
H_{x_t} = \frac{E_t}{2} \bigl[ \cos (\theta_t) \sigma_z + \sin (\theta_t) \sigma_x \bigr] ~,
\end{align}
where $\sigma_x$, $\sigma_y$, and $\sigma_z$ are the Pauli operators.
While $E_t$ quantifies the energy gap between the system's instantaneous energy eigenstates,
$\theta_t$ parametrizes the instantaneous orientation of the energy eigenbasis relative to the `computational' $z$-basis.
If we assume the Markovian limit and detailed-balanced dynamics as the system interacts weakly with a large bosonic bath, then
the system evolves according to the instantaneous Lindbladian~\cite{Mill20}:
\begin{align}
\dot{\rho}_t = \mathcal{L}_{x_t} (\rho_t)
= \frac{i}{\hbar} [\rho_t, H_{x_t}] 
& + \frac{c E_t}{\hbar} (N_{x_t} + 1) \mathcal{D}[L_{x_t}] (\rho_t) \nonumber \\
& + \frac{c E_t}{\hbar} N_{x_t} \mathcal{D}[L_{x_t}^\dagger] (\rho_t)
\end{align} 
where $\mathcal{D}[L](\rho) = L \rho L^\dagger - \tfrac{1}{2} \{ L^\dagger L, \rho \}$,
$N_{x_t} = (e^{\beta E_t} - 1)^{-1}$,
and $c$ is the coupling strength to the bath.
The time-dependent lowering operator can be represented as
\begin{align*}
L_{x_t} = \frac{1}{2} 
\bigl[
\cos (\theta_t) \sigma_x
- i \sigma_y
- \sin (\theta_t) \sigma_z  
\bigr]
\end{align*}
and satisfies the detailed balance condition 
$[L_{x_t}, H_{x_t}] = E_t L_{x_t}$ ~\cite{Mill20, Manz15}.

This generic Hamiltonian and Lindbladian have a long history of applications to many distinct physical systems~\cite{Legg87},
and have been used recently to explore the nonequilibrium thermodynamics of qubit erasure~\cite{Mill20}.

We consider three control protocols
to reset the system to the computational-basis state $\ket{0} = \sigma_z \ket{0}$.
For all protocols, 
we choose the coupling strength $c = 1/5$, the duration $\tau = 50 \beta \hbar$, and the time-step for numerical integration $dt = \beta \hbar / 500$.

\begin{figure}[H]
\begin{center}
\includegraphics[width=0.95\columnwidth]{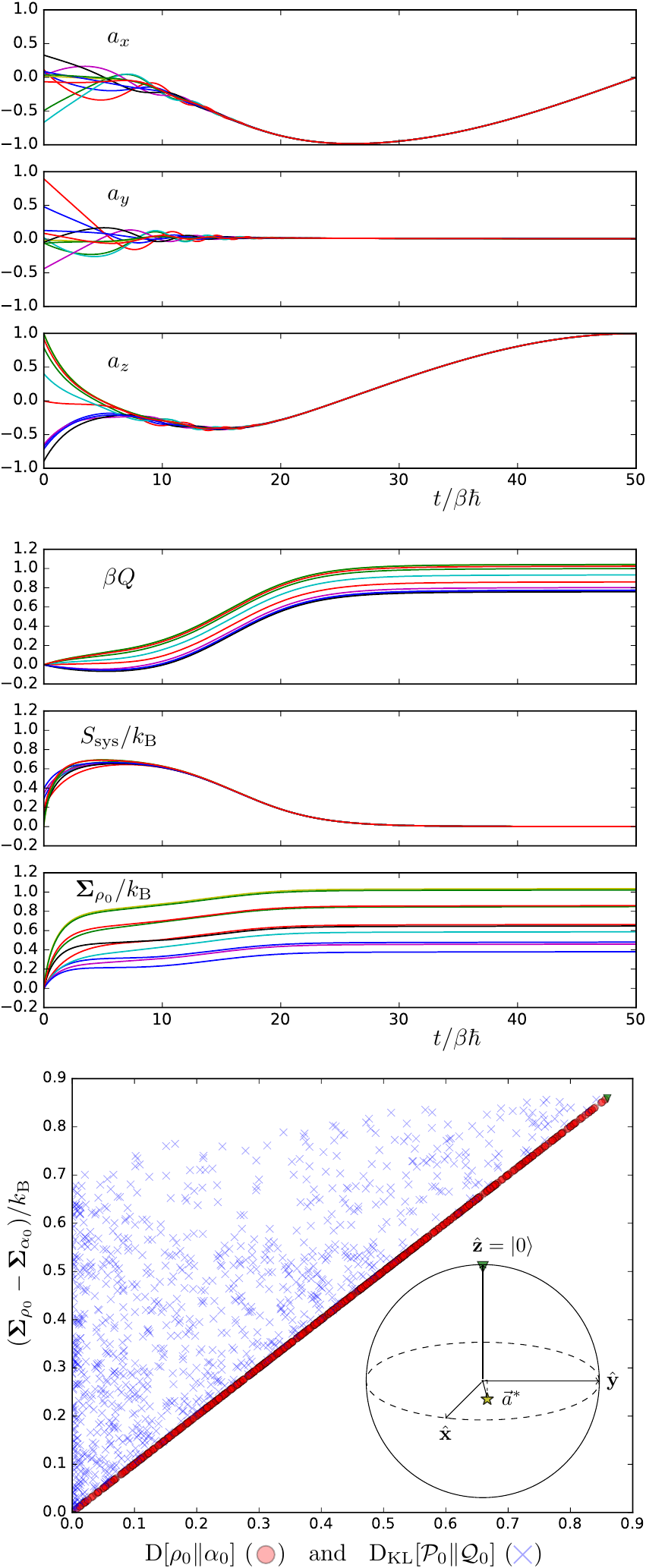}
\end{center}
\caption{A reliable protocol for erasing a qubit,
which evolves the energy gap and orientation of the 
energy eigenstates via $E_t = E_0 + (E_\tau - E_0) \sin^2( \pi t / 2 \tau)$
and $\theta_t = \pi t / \tau$ respectively.
Top three panels:  
Evolution of the Bloch vector 
for ten initial states sampled randomly from the Bloch sphere.
Next three panels:
Evolution of thermodynamic quantities for the same ten initial states.
Bottom panel:
Dissipation for 1000 random initial states, compared to the 
minimally dissipative state.
	}
\label{fig:QuantumTAP1}
\end{figure}

\begin{figure}[H]
\begin{center}
\includegraphics[width=0.95\columnwidth]{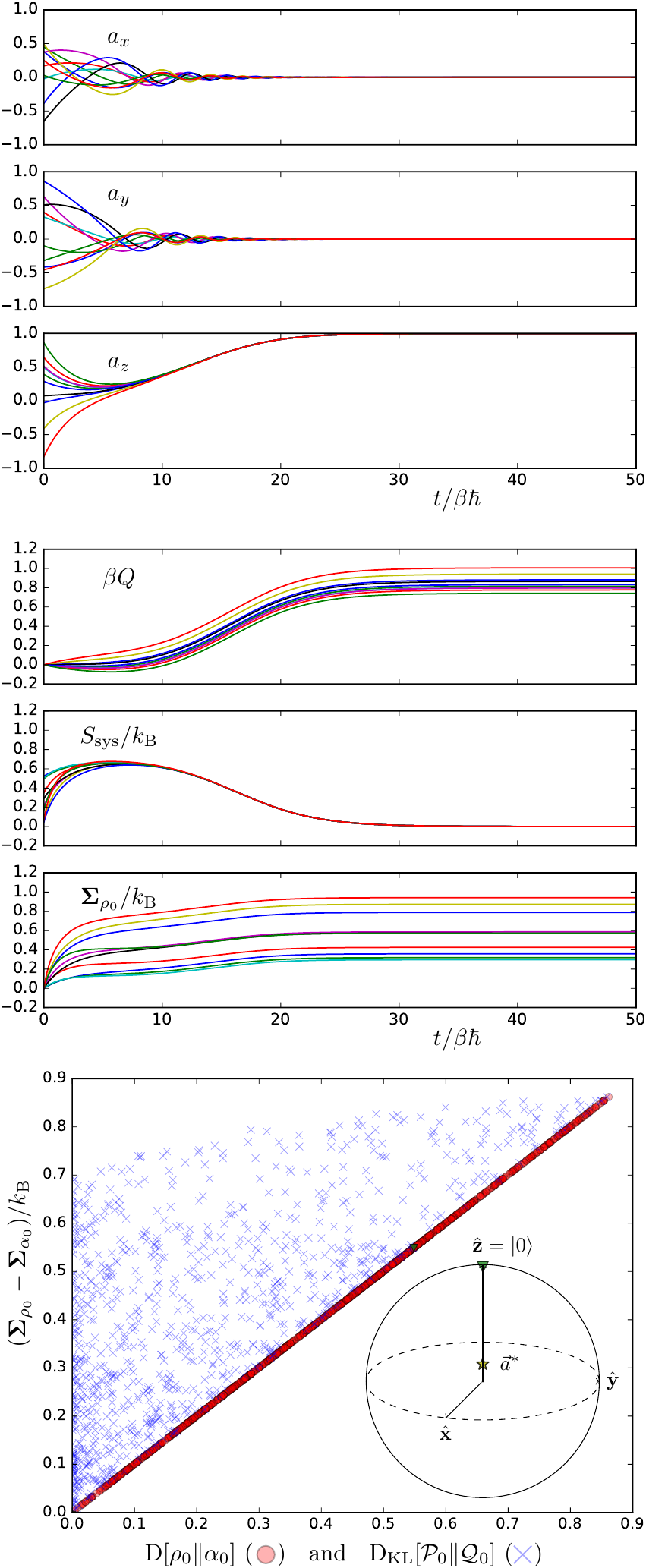}
\end{center}
\caption{A reliable protocol for erasing a qubit,
which evolves the energy gap via $E_t = E_0 + (E_\tau - E_0) \sin^2( \pi t / 2 \tau)$
while the orientation of the 
energy eigenstates is fixed via
$\theta_t = \pi$.
Top three panels:  
Evolution of the Bloch vector 
for ten initial states sampled randomly from the Bloch sphere.
Next three panels:
Evolution of thermodynamic quantities for the same ten initial states.
Bottom panel:
Dissipation for 1000 random initial states, compared to the 
minimally dissipative state.
	}
\label{fig:QuantumTAP2}
\end{figure}

For the first two protocols, the  gap between energy eigenvalues changes smoothly as
$E_t = E_0 + (E_\tau - E_0) \sin^2(\frac{\pi t}{ 2 \tau})$ with $E_\tau = 10 \kB T$ and $E_0 = E_\tau / 50$.
During the protocol
analyzed in Fig.~\ref{fig:QuantumTAP1},
the energy eigenstates rotate
according to
$\theta_t = \pi t / \tau$.
Whereas, 
during the alternative protocol
analyzed in Fig.~\ref{fig:QuantumTAP2},
the energy eigenstates have a fixed orientation
according to
$\theta_t = \pi$.
These first two protocols are very similar to Ref.~\cite{Mill20}, but with slightly 
different parameter settings, and analyzed for different purposes.

\begin{figure}[H]
\begin{center}
\includegraphics[width=0.95\columnwidth]{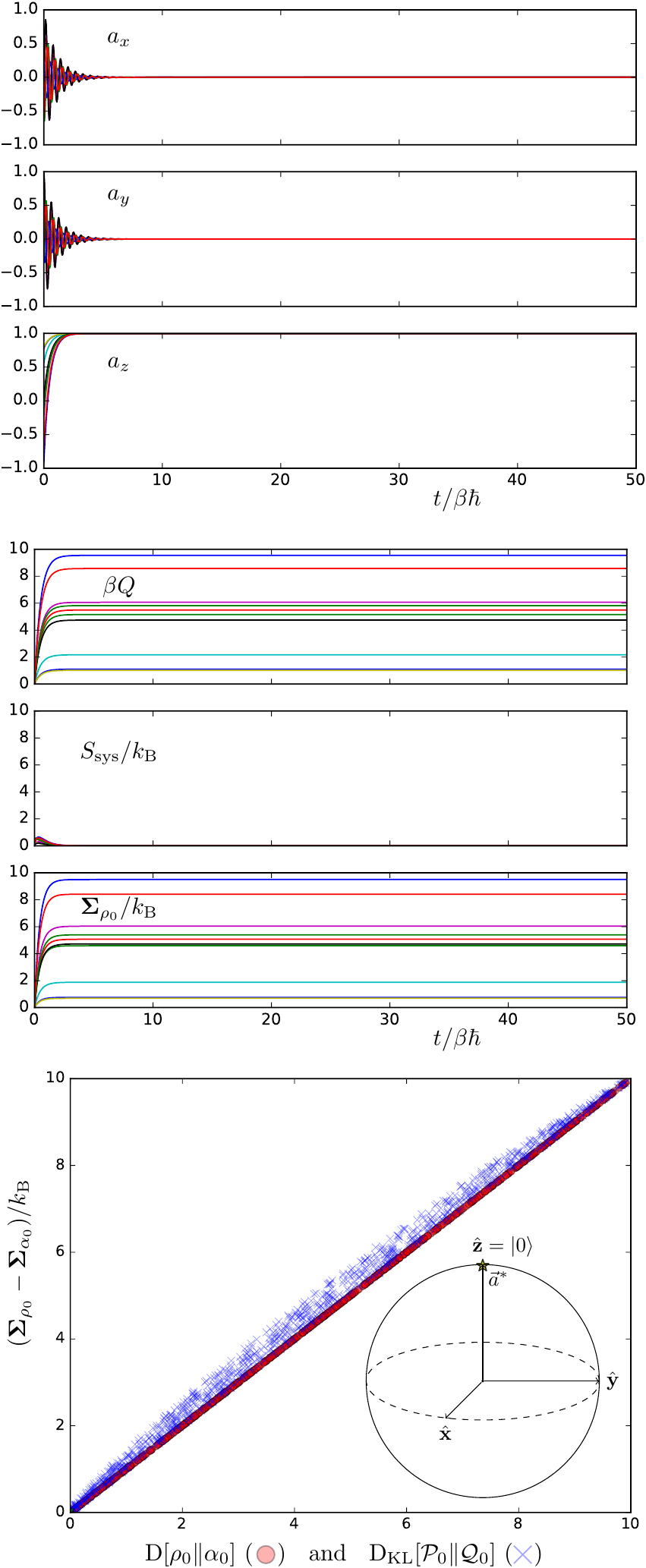}
\end{center}
\caption{A reliable protocol for erasing a qubit,
which achieves reset via relaxation to 
thermal equilibrium 
via fixed energy eigenstates with 
$E_t = 10 \kB T$ 
and $\theta_t = \pi$ for $0 < t \leq \tau$.
Top three panels:  
Evolution of the Bloch vector 
for ten initial states sampled randomly from the Bloch sphere.
Next three panels:
Evolution of thermodynamic quantities for the same ten initial states.
Bottom panel:
Dissipation for 1000 random initial states, compared to the 
minimally dissipative state.
	}
\label{fig:QuantumTAP3}
\end{figure}

The final protocol, shown in Fig.~\ref{fig:QuantumTAP3},
achieves reset simply via relaxation to an equilibrium state which (via a large energy gap of $10 \kB T$) 
is designed to be very close to a pure state.

In each Fig.~\ref{fig:QuantumTAP1}, \ref{fig:QuantumTAP2}, and \ref{fig:QuantumTAP3},
we show the evolution of ten randomly sampled initial density matrices via the evolution of their Bloch vectors. 
These are the first three panels of each figure.
The next three panels of each figure show the evolution of expected heat, system entropy, and expected entropy production
that ensues from each of these ten randomly sampled initial density matrices.
The wide distribution in expected heat and expected entropy production is evident
for each protocol.

For each protocol, there is a single
minimally dissipative initial state.
Our examples demonstrate that, in general, 
this is neither the initial equilibrium state nor the desired final state.
This is most apparent for the first protocol, 
since the minimally dissipative state is not diagonalized in the computational basis that diagonalizes 
both the initial equilibrium state and the desired reset state.  
(To see this, note that $\vec{a}^*$ is not along the $\hat{\textbf{z}}$-axis 
in the bottom panel of Fig.~\ref{fig:QuantumTAP1}.)

For each quantum erasure protocol,
we find the unique minimally dissipative initial density matrix $\q_0$
via Eqs.~\eqref{eq:EFvector_solution} and \eqref{eq:Final_avec_condition}---from
the entropy flow observed from four of the random initial conditions.
In the bottom panel of each figure,
dissipation vs.\ the quantum relative entropy D$[ \rho_0 \| \q_0 ]$ is shown in red circles for 
1000 random initial states.
The red circles all lie along the identity, which clearly verifies Eq.~\eqref{eq:MainResetEqForEP}.

In the same bottom panels, 
we show the position of $\q_0$ in the Bloch sphere via its Bloch vector $\vec{a}^*$.
Furthermore, 
we exhibit the thermodynamic role of coherence via the position of the blue crosses.
Via Eq.~\eqref{eq:RelEnt_asDKLplusCoherence},
the horizontal gap between D$[\rho_0 \| \q_0]$ (red circles) and 
D$_\text{KL}[ \mathcal{P}_0 \| \mathcal{Q}_0 ]$ (blue crosses)
quantifies the initial coherence $C_{\q_0}(\rho_0)$.
Finally, 
the green triangle along the identity shows
the dissipation incurred when the initial state is the desired final state.
Starting where you end is not typically optimal; indeed,
Fig.~\ref{fig:QuantumTAP1} shows that this is nearly the worst way to start for that protocol.

While the distribution of thermodynamic quantities
appears similar for the first two protocols, 
we emphasize
that the protocols are efficient for different initial states. 	
This is seen most immediately via the bottom panel of Figs.~\ref{fig:QuantumTAP1}	and \ref{fig:QuantumTAP2},
where it is apparent that each protocol has a distinct minimally dissipative state.
States closest to the respective minimally dissipative states are efficient for the respective protocols.

The typical dissipation in Fig.~\ref{fig:QuantumTAP3} is much larger than in the previous two cases, since the 
minimally dissipative state is closer to a pure state.
In this case, it is interesting to note that
the Kullback--Leibler divergence is responsible for most of the dissipation,
with coherence playing a relatively minor role.
Recall that the relative entropy of coherence is always upper bounded by a constant, $\ln d$, 
for a $d$-dimensional system.
Hence, 
since this reset process is capable of very large entropy production $\EP \gg \ln d = \ln 2$,
the classical limit explains the bulk of the dissipation.

\section{Conclusion}

Landauer's bound for the heat required to reset a state
is achieved in the limit of 
zero entropy production.
However, we have shown that bound to be unachievable 
for almost every quantum input to any thermodynamic process for reliable reset.
This follows from the fact that there is a unique initial quantum state leading to minimal entropy production for each reliable reset protocol.
Moreover, we have demonstrated that when a reset protocol is modified to bring this minimally dissipative initial state closer to a pure state, entropy production (and thus also heat) diverges for all inputs
that are not close to this pure state.

For the case of qubit reset, we have found the minimally dissipative state analytically in terms of the entropy-flow vector, 
which was introduced here.
We anticipate that the entropy-flow vector may be a useful concept also for larger quantum systems, via a straightforward generalization of Eq.~\eqref{eq:EF_simpleLinAlg}.

Landauer's bound is often invoked to draw physical implications from quantum information theory, via the universality of thermodynamics.  
Our results show that this connection is somewhat tenuous, since Landauer's bound is impossible to achieve for almost every initial state for any reliable reset mechanism.  Nevertheless, nonequilibrium thermodynamics offers refined information-theoretic equalities,
like Eq.~\eqref{eq:QKW_result}, that are valid and offer tight predictions arbitrarily far from equilibrium.
Recognizing the impossibility of Landauer's bound will be especially important for designing practical quantum computers, in which quantum states must be reliably prepared while maintaining coherence through low-heat transformations.

The classical version of these results answer another interesting and non-trivial problem, since the computational states are then typically collective metastable states.  We address the impossibility of the classical Landauer bound in Ref.~\cite{Riec20b}.

\acknowledgements

This work was supported by the
National Research Foundation and
L'Agence Nationale de la Recherche joint Project No.~NRF2017-NRFANR004 VanQuTe,
the National Research Foundation of Singapore Fellowship No.~NRF-NRFF2016-02,
the Singapore Ministry of Education Tier 1 grant RG190/17,
and the FQXi Grant `Are quantum agents more energetically efficient at making predictions?'.

\appendix

\section{Entropy Flow}
\label{sec:EF_def}

We find that the expected entropy flow can most generally be 
represented as
\begin{align}
\EF \equiv - \kB \int_0^\tau \tr(\dot{\rho}_t^\text{ env} \ln \stationary_{x_t}^\text{env}) \, dt ~,
\label{eq:GenEFdef}
\end{align}
where 
$\rho_t^\text{ env}$ is the reduced state of the environment at time $t$.
$\stationary_{x_t}^\text{env}$ is a time-dependent 
reference state
that represents the environment as a set of thermodynamic baths
$\mathbb{B}$ in local equilibrium:
$\stationary_{x_t}^\text{env} = \bigotimes_{b \in \mathbb{B}} \stationary_{x_t}^{(b)}$.
The equilibrium state $\stationary_{x_t}^{(b)}$
is constructed 
with the bath's operators (e.g., Hamiltonian $H_{x_t}^{(b)}$, number operators $N_{x_t}^{(b, \ell)}$, etc.) that correspond to its variable observable quantities (energy, particle numbers, etc.)~\cite{Reic09, Albe01}.

For example, if the temperatures $\{ T_t^{(b)} \}_{b \in \mathbb{B}}$
of the baths are varied over the course of a protocol, and the baths can only exchange energy,
then
the equilibrium states of the baths
are canonical
and
Eq.~\eqref{eq:GenEFdef} reduces to the familiar expression: 
\begin{align}
\EF = \sum_{b \in \mathbb{B}} \int 
\frac{\delta Q^{(b)}}{T_t^{(b)}} ~,
\end{align}
where $\delta Q^{(b)}$ is a small transfer of energy to bath $b$.

As another example, if each bath has a time-independent grand canonical reference  
state with temperature $T^{(b)}$ and 
chemical potentials $\{ \mu^{(b, \ell)} \}_\ell$,
then 
Eq.~\eqref{eq:GenEFdef} reduces to
Eq.~\eqref{eq:SenvDef1} of the main text:
\begin{align*}
\EF
= \sum_{b \in \mathbb{B}} \frac{Q^{(b)}}{T^{(b)}} - \frac{1}{T^{(b)}} \sum_{\ell} \mu^{(b, \ell)} \Delta \braket{N^{(b, \ell)}} ~.
\end{align*}
Eq.~\eqref{eq:SenvDef1} has been used to 
explore entropy production even in the case of arbitrarily small baths, by fixing the reference state via the initial temperature and chemical potentials~\cite{Espo10a, Ptas19a}.

\section{Initial-state dependence of entropy production}
\label{sec:EP_InitStateDependence}

Entropy production is defined by Eq.~\eqref{eq:EPdef} of the main text, 
where the entropy flow is most generally defined by Eq.~\eqref{eq:GenEFdef}:
\begin{align*}
\EF \equiv - \kB \int_0^\tau \tr(\dot{\rho}_t^\text{ env} \ln \stationary_{x_t}^\text{env}) \, dt ~.
\end{align*}
The environmental reference state $\stationary_{x_t}^\text{env}$ in Eq.~\eqref{eq:GenEFdef} 
may be time-dependent, but is assumed to be independent of the initial state of the system.
Under this assumption,
the expected entropy flow is 
a linear function of the initial state of the system. This linearity is evident once we write out the 
reduced state of the environment as:
$\rho_t^\text{env} = \trsys(U_{t} \rho_0 \otimes \rho_0^\text{env} U_t^\dagger )$. 
Because 
the expected entropy flow is 
a linear function of the initial state of the system, 
Thm.~2 of
Ref.~\cite{Riec20a} guarantees that the expected entropy production from any initial state $\rho_0$ is quantified by:
\begin{align}
\EP - \EP[\alpha_0] = \kB \text{D} [ \rho_0 \| \alpha_0 ] - \kB \text{D} [ \rho_\tau \| \alpha_\tau ] ~,
\label{eq:MainEqualityInApp}
\end{align}
where 
$$ \alpha_0 \in \argmin_{\rho_0} \EP~. $$
This generalizes the main result of Ref.~\cite{Riec20a} to allow for a time-dependent environmental reference state.
This establishes, for example, that  Eq.~\eqref{eq:MainEqualityInApp} remains valid in scenarios where
 the temperature profile of a bath may be deterministically modulated through time.

\begin{widetext}

\section{Any reliable reset operation has a unique minimally dissipative state with full support}
\label{sec:q0HasFullSupport}

We will prove that the minimally dissipative input state to any reliable reset protocol 
will have full support on the domain of states to be reset.

Consider an initial state $\xi_0$ that does \emph{not} have full support.
This state has a collection of eigenvalues $\Lambda$
and a spectral decomposition $\xi_0 = \sum_{\lambda \in \Lambda} \lambda \ket{\lambda} \! \bra{\lambda}$
with $0 \leq \lambda \leq 1$ for all $\lambda$ and $\sum_{\lambda \in \Lambda} \lambda = 1$.
By assumption, $\xi_0$ must have a spectral decomposition with at least one zero eigenvalue
with associated spectral projection operator $\ket{ o } \! \bra{ o }$
that acts on a subspace of the domain to be reset.
(We use $\ket{ o }$ here for the eigenstate associated with the eigenvalue of zero, 
so that it is not confused with a computational-basis state.)

We will now show that---if entropy flow is finite for all inputs---there 
is always a state $\xi_0' = (1-\delta) \xi_0 + \delta \ket{ o } \! \bra{ o }$
with greater support than $\xi_0$ that dissipates less than $\xi_0$ for some $0 < \delta < 1$. 
Accordingly, a state that lacks full support will never be the minimally dissipative input to a reliable reset protocol.

Recall from Eq.~\eqref{eq:EPdef_forReliableReset} that the entropy production during any reliable reset protocol is
\begin{align*}
\EP = \EF  + \kB S(r_\tau) - \kB S(\rho_0) ~,
\end{align*}
where $r_\tau$ is the input-independent final state.
Furthermore,
recall from Sec.~\ref{sec:EP_InitStateDependence} that the expected entropy flow is a linear function of the initial state.
Note that the eigenvalues of $\xi_0'$ are $(1-\delta) \lambda$ for each $\lambda \in \Lambda \setminus \{ 0 \}$, while one of the zero-eigenvalues of $\xi_0$ maps to an eigenvalue of $\delta$ for $\xi_0'$. 

The entropy production from input $\xi_0'$ is thus:
\begin{align}
\EP[\xi_0'] 
&=
\EP[(1-\delta) \xi_0 + \delta \ket{ o } \! \bra{ o }] \\
&= \delta \EF[ \ket{ o } \! \bra{ o } ] + (1-\delta) \EF[\xi_0] + \kB \bigl\{ (1-\delta) \sum_{\lambda \in \Lambda} \lambda \ln [ \lambda (1 - \delta)] \bigr\} + \kB \delta \ln \delta + \kB S(r_\tau) \\
&= \delta \EF[ \ket{ o } \! \bra{ o } ] + (1-\delta) \EF[\xi_0] -  \kB (1-\delta) S(\xi_0) + \kB  (1-\delta) \ln  (1-\delta)
+ \kB \delta \ln \delta + \kB S(r_\tau) \\
&= \delta \EP[ \ket{ o } \! \bra{ o } ] + (1- \delta) \EP[ \xi_0 ] - \kB \text{B}(\delta) ~,
\label{eq:EPfromConvexMixture}
\end{align}
where B$(\delta) = - \delta \ln \delta -  (1-\delta) \ln  (1-\delta)$ is the binary entropy function,
and we have used the fact that 
$\sum_{\lambda \in \Lambda} \lambda = 1$.

If the difference between $\EP[\xi_0] $ and $\EP[\xi_0'] $ is positive, then $\xi_0$ cannot be a minimally dissipative state.
From Eq.~\eqref{eq:EPfromConvexMixture},
we find:
\begin{align}
\EP[\xi_0]  -  \EP[\xi_0'] 
&=  \kB \text{B}(\delta) 
- \delta  ( \EP[ \ket{ o } \! \bra{ o } ] -  \EP[ \xi_0 ] )
~.
\end{align}
If $\delta = 0$ then, of course,
$\EP[\xi_0]  = \EP[\xi_0'] $ since $\xi_0'$ is then in fact equal to $\xi_0$.
However, we find that $\EP[\xi_0]  - \EP[\xi_0'] $ has a positive derivative with respect to $\delta$,
from $\delta=0$ up to some sufficiently small finite value.
I.e., 
\begin{align}
\tfrac{d}{d \delta} (
\EP[\xi_0]  -  \EP[\xi_0'] )
&=  \kB \ln \Bigl( \frac{1-\delta}{\delta} \Bigr)
-  ( \EP[ \ket{ o } \! \bra{ o } ] -  \EP[ \xi_0 ] )
~.
\end{align}
Since $\EP[ \ket{ o } \! \bra{ o } ] -  \EP[ \xi_0 ] $ is assumed to be 
finite---while $\ln \bigl( \frac{1-\delta}{\delta} \bigr)$ grows unbounded 
as $\delta \to 0$---this tells us that $\EP[\xi_0]  > \EP[\xi_0'] $ for sufficiently small $\delta$.
Hence, $\xi_0$ cannot be the minimally dissipative state of any reliable reset protocol.
Notably, the only assumption that we made about $\xi_0$ is that it does not have full support.

We have shown that any state lacking full support cannot be the minimally dissipative input for any reliable reset protocol.
Thus, the minimally dissipative state $\q_0$ must have full support on the domain of states to be reset.

By Eq.~\eqref{eq:MainResetEqForEP},
the dissipation from any input state $\rho_0$ 
exceeds this minimal dissipation by $\kB \text{D} [\rho_0 \| \q_0]$,
which is positive for all $\rho_0 \neq \q_0$.

\section{Existence of---and an Expression for---the Entropy Flow Vector}
\label{sec:ExistenceAndConstruction_for_EFvec}

Entropy flow is an affine function of the initial state.
Recall from Eq.~\eqref{eq:GenEFdef}
that the expected
entropy flow can be expressed as:
\begin{align}
\EF 
&= - \kB \int_0^\tau \tr(\dot{\rho}_t^\text{ env} \ln \stationary_{x_t}^\text{env}) \, dt \nonumber \\
&= -  \kB \int_0^\tau \tr \Bigl\{ \bigl[ \tfrac{d}{dt} \tr_\text{sys} ( U_{x_{0:t}} \rho_0 \otimes \rho_0^\text{ env} U_{x_{0:t}}^\dagger ) \bigr]  \ln \stationary_{x_t}^\text{env} \Bigr\} \, dt 
~,
\label{eq:EF_suitable_for_derivation}
\end{align}
where the environmental reference state $\stationary_{x_t}^\text{env}$ is independent of the initial state of the system.

For any initial state $\rho_0 = I/2 + \tfrac{1}{2} \vec{a} \cdot \vec{\sigma}$,
it is then clear that
\begin{align}
\EF
&=
- \kB \int_0^\tau \tr \Bigl( \Bigl\{ \tfrac{d}{dt} \tr_\text{sys} \bigl[ U_{x_{0:t}} ( I/2 + \tfrac{1}{2} \vec{a} \cdot \vec{\sigma} ) \otimes \rho_0^\text{ env} U_{x_{0:t}}^\dagger \bigr] \Bigr\}  \ln \stationary_{x_t}^\text{env} \Bigr) \, dt \\
&= 
\EF[I/2]
- \tfrac{1}{2} \vec{a} \cdot
\kB \int_0^\tau \tr \Bigl\{ \bigl[ \tfrac{d}{dt} \tr_\text{sys} ( U_{x_{0:t}} \vec{\sigma} \otimes \rho_0^\text{ env} U_{x_{0:t}}^\dagger ) \bigr]  \ln \stationary_{x_t}^\text{env} \Bigr\} \, dt ~.
\label{eq:EF_suitable_for_derivation2}
\end{align}

This shows that there exists an initial-state-independent vector $\vec{\phi} = (\phi_x , \, \phi_y , \, \phi_z)$ such that
\begin{align}
\EF = \EF[I/2] +  \tfrac{1}{2} \vec{a} \cdot \vec{\phi} ~.
\label{eq:EF_simpleLinAlg_App}
\end{align}
We call $\vec{\phi}$ the ``entropy-flow vector'' induced by the control protocol.

Beyond merely proving its existence, Eq.~\eqref{eq:EF_suitable_for_derivation2} gives an explicit construction 
for the entropy-flow vector:
\begin{align}
\vec{\phi} = 
- \kB \int_0^\tau \tr \Bigl\{ \bigl[ \tfrac{d}{dt} \tr_\text{sys} ( U_{x_{0:t}} \vec{\sigma} \otimes \rho_0^\text{ env} U_{x_{0:t}}^\dagger ) \bigr]  \ln \stationary_{x_t}^\text{env} \Bigr\} \, dt 
        ~.
\label{eq:EFvec_construction}
\end{align}
Alternatively, this could be written component-wise as:
\begin{align}
\phi_k = 
- \kB \int_0^\tau \tr \Bigl\{ \bigl[ \tfrac{d}{dt} \tr_\text{sys} ( U_{x_{0:t}} \sigma_k \otimes \rho_0^\text{ env} U_{x_{0:t}}^\dagger ) \bigr]  \ln \stationary_{x_t}^\text{env} \Bigr\} \, dt 
\label{eq:EFvec_componentwise_construction}
\end{align}
for $k \in \{ x, y, z \}$.

When the environmental reference state is time independent, such that $\stationary_{x_t}^\text{env} =  \bigotimes_{b \in \mathbb{B}} \stationary^{(b)} $,
this reduces to
\begin{align}
\vec{\phi} = 
- 
\sum_{b \in \mathbb{B}} \kB  \tr \bigl[ \trallbutb ( \mathcal{U}_{\drive}   \vec{\sigma}  \otimes \rho_0^\text{env}  \mathcal{U}_{\drive}^\dagger ) \ln  \stationary^{(b)} \bigr]             
        ~.
\label{eq:EF_special_vec_construction}
\end{align}

\end{widetext}

\section{Further Implementations for Quantum Memory Reset in the Weak Coupling Regime}

Here we discuss an approach to memory reset in the weak coupling regime.

We first consider the simplest scenario of single qubit reset to $\ket{0}$.
An obvious way to implement reset is the following: 
\begin{enumerate}
\item
Attach the qubit to a single thermal bath at temperature $T$.
\item
Change the system Hamiltonian from $H_{x_0}$ to $H_{x_\tau} = E \ket{1}\bra{1}$ with $E \gg \kB T$.
\item
Allow the system to equilibrate, so that $\rho_\tau \approx \stationary_{x_\tau} \approx \ket{0} \bra{0}$.
\end{enumerate}
If the desired reset state is a pure state $r_\tau \neq \ket{0} \bra{0}$, then the system can subsequently be detached from the 
thermal reservoir and unitarily evolved to $r_\tau$.
Alternatively, the desired state can be attained more directly by driving to the final Hamiltonian $H_{x_\tau} = E ( I - r_\tau)$.
With the luxury of much time, the error can be made arbitrarily small.

A related procedure allows reliable reset to any state $r_\tau$:
\begin{enumerate}
\item
Attach the qubit to a single thermal bath at temperature $T$.
\item
Change the system Hamiltonian from $H_{x_0}$ to $H_{x_\tau} = - \kB T \ln r_\tau + c$, where $c$ is an arbitrary constant.
\item
Allow the system to equilibrate to $\stationary_{x_\tau} = r_\tau$.
\end{enumerate}

In the quasistatic limit, this reset can be achieved with zero entropy production \emph{if} the initial state is known.
The protocol can then be designed to make $\rho_0 = \q_0$.
To achieve zero dissipation, the initial system-Hamiltonian should change instantaneously
from $H_{x_0}$ to $H_{x_{0^+}} = - \kB T \ln \rho_0  + c'$, where $c'$ is an arbitrary constant.
Subsequently, the Hamiltonian should be changed quasistatically (via any continuous path) 
from $H_{x_{0^+}}$ to $H_{x_\tau}$.
In such scenarios, the work invested will equal the change in free energy,
so that there is zero entropy production.
If, however, the initial state $\rho_0$ is \emph{not} equal to the anticipated state $\q_0 = \pi_{x_{0^+}} \propto e^{- \beta H_{x_{0^+}} }$,
then there will be dissipation equal to the loss of initially induced 
nonequilibrium addition to free energy: 
$\EP = \kB \text{D}[ \rho_0 \|  \pi_{x_{0^+}} ]$.

In the slow but finite-duration linear response regime, 
reliable reset can still be achieved within some small error tolerance $\epsilon$,
but the dissipation will scale as $1/\tau$.
Optimal protocols in the linear response regime can be found as natural geodesics
induced by the friction tensor~\cite{Siva12, Mand16, Scan19}.

As the protocol becomes quick relative to the relaxation timescales for the system,
the protocol must be modified to maintain reliable reset to the desired state.
One way to achieve this is via counter-diabatic alterations to the driving~\cite{Vaca14}.
Counter-diabatic driving can, for example,
enforce that the system stays along the optimal linear-response state trajectory.

In these finite-time scenarios, the dissipation will be nontrivial for all initial states, 
but our equality relating dissipation among all possible initial states will remain valid.

These results can be easily extended to a scenario for quantum reset of $N$ qubits.
An obvious way to implement reset is the following: 
\begin{enumerate}
\item
Attach the $N$ qubits to a single thermal bath at temperature $T$.
\item
Change the system Hamiltonian from $H_{x_0}$ to $H_{x_\tau} = E \bigl( I - (\ket{0} \bra{0})^{\otimes N} \bigr)$ with $E \gg \kB T$.
\item
Allow the system to equilibrate, so that $\rho_\tau \approx \stationary_{x_\tau} \approx (\ket{0} \bra{0})^{\otimes N}$.
\item
Detach the qubits from the thermal bath 
and unitarily transform them to any desired $N$-qubit pure state.
\end{enumerate}
This allows, for example, reset to the maximally entangled state.
Alternatively, the desired reset state could be achieved more directly
by designing the final Hamiltonian $H_{x_\tau}$ such that $r_\tau$ is its equilibrium state.
This suggests the alternative procedure for resetting to any $N$-qubit state:
\begin{enumerate}
\item
Attach the $N$ qubits to a single thermal bath at temperature $T$.
\item
Change the system Hamiltonian from $H_{x_0}$ to $H_{x_\tau} = - \kB T \ln r_\tau + c$, where $c$ is an arbitrary constant.
\item
Allow the system to equilibrate to $\stationary_{x_\tau} = r_\tau$.
\end{enumerate}

The earlier comments about finite-time protocols---linear response, counter-diabatic driving, etc.---all apply 
to this general $N$-qubit reset as well.

\end{document}